\documentclass[useAMS,usenatbib]{mn2e}

\usepackage{lscape}
\usepackage{graphicx}
\usepackage{natbib}
\usepackage{amsmath,amssymb}
\newcommand{\halpha}{H{$\alpha$}}
\newcommand{\hbeta}{H{$\beta$}}

\newcommand{\loiii}{$L_{{\rm [O\,III]}}$}

\newcommand{\NII}{[N{\sevenrm\,II}]}
\newcommand{\NIIb}{[N{\sevenrm\,II}]\,$\lambda$6584}
\newcommand{\OIII}{[O{\sevenrm\,III}]}

\newcommand{\OIIIb}{[O{\sevenrm\,III}]\,$\lambda$5007}

\newcommand{\ssfr}{SSFR$_{\tau}$}

 \font\sevenrm=cmr7 scaled 1000

\begin{document}

\title[BULGE STAR FORMATION HISTORY IN AGNs]{Tracing the History of Recent
Bulge Star Formation in Active Galactic Nuclei}

\author[X. LIU]{Xin Liu$^{1}$\thanks{E-mail: xinliu@astro.princeton.edu} \\
$^{1}$Department of Astrophysical Sciences, Princeton University,
Peyton Hall -- Ivy Lane, Princeton, NJ 08544}

\date{Accepted 2010 May 12. Received 2010 May 11; in original form 2009 November 25}

\maketitle


\begin{abstract}
We examine the relation between black hole accretion and bulge
star formation as a function of look-back time ($\tau$) in 20,541
obscured AGNs (with redshifts $\bar{z}\sim$0.1 and bolometric
luminosities $L_{{\rm Bol}}\sim10^{43}$--$10^{45}$ erg s$^{-1}$)
optically selected from the Sloan Digital Sky Survey (SDSS). To
quantify the most recently formed stars with ages less than
typical AGN lifetimes, we estimate the differentiated specific
star formation rate (SSFR$_{\tau}$) based on population synthesis
analysis.  Eddington ratio ($\lambda$) is inferred using \OIIIb\
luminosity and stellar velocity dispersion as proxies for $L_{{\rm
Bol}}$ and black hole mass respectively. We find that when $\tau <
\tau_{0}$, SDSS AGNs follow a power law $\lambda \propto$
SSFR$_{\tau}^{1.0-1.1}$; the relation flattens out when $\tau >
\tau_{0}$. The threshold timescale $\tau_{0}$ is $\sim 0.1$ ($\sim
1$) Gyr in young (old) bulges. The scatter in the power laws is
dominated by observational uncertainties. These results may
provide useful constraints on models explaining the correlations
between AGN activity and bulge star formation.
\end{abstract}


\section{Introduction}\label{sec:intro}

The growth of supermassive black holes (SMBHs) could be linked
with star formation (SF) in their host-galaxy bulges through
multiple processes \citep[see e.g.,][for a review]{heckman08}.
Stars form under the cosmological infall of gas
\citep{binney77,rees77}, some of which could be accreted by
central SMBHs \citep[e.g.,][]{burkert01}.  Galaxy mergers or
interactions may trigger starburst and quasar activity
\citep[e.g.,][]{sanders88,moore96}, which could output
energy and momentum regulating further growth
\citep[e.g.,][]{silk98,fabian99,springel05}.  AGN-driven outflows
could boost starbursts \citep[e.g.,][]{begelman89,silk09}. After
being active, evolved stars may fuel AGNs via mass losses
\citep[e.g.,][]{norman88,ciotti97}.  These processes may have
different significance in regulating SMBH growth and bulge SF in
various AGN populations at different cosmic epochs.

While post-starburst populations have been extensively observed in
quasars \citep[e.g.,][]{boroson82,canalizo00,jahnke07,liu09} and
in higher-luminosity AGNs
\citep[e.g.,][]{kauffmann03,heckman04,cid04,netzer09}, the
physical connection is still in debate.  Two useful parameters are
specific star formation rate (${\rm SSFR} \equiv {\rm
SFR}/M_{\ast}$) and Eddington ratio ($\lambda \equiv L_{{\rm
Bol}}/L_{{\rm Edd}}$).  By studying the $\lambda$ distribution in
AGNs as a function of mean stellar age, \citet{kauffmann09}
suggested that local AGNs occupy two distinct growth regimes.
These authors used the 4000-{\rm \AA} break
\citep[D$_n(4000)$;][]{bruzual83,balogh99} to characterize mean
stellar age.  Arising from a series of metallic lines, D$_n(4000)$
is most sensitive to 1--2 Gyr old stars \citep[e.g.,][]{bc03},
whereas luminous AGN phases appear to be brief \citep[$<$0.1 Gyr;
e.g.,][]{yu02}.

To get higher temporal resolution and to quantify the most
recently formed stars with ages less than typical AGN lifetimes,
we examine SSFR averaged in the past $\tau$ years (SSFR$_{\tau}$)
based on population synthesis analysis
\citep[e.g.,][]{bica88,cid05,asari07}.  Under the
multiple-starburst approximation, the differentiated SF histories
in a statistical sample of AGNs offered by the Sloan Digital Sky
Survey \citep[SDSS;][]{york00} enable us to investigate the
correlation between AGN and recent bulge SF activity as a function
of look-back time.  This approach offers insights otherwise
unattainable with single and older age indicators.  We describe the
AGN sample and the assumptions in estimating $\lambda$ and
SSFR$_{\tau}$ along with uncertainties and selection biases in \S
\ref{sec:sample}. \S \ref{sec:corr} focuses on the
correlations between $\lambda$ and SSFR$_{\tau}$ as a function of
$\tau$.  We discuss implications and caveats in \S
\ref{sec:discuss}.  A cosmology with $\Omega_m = 0.3$,
$\Omega_{\Lambda} = 0.7$, and $h = 0.7$ is assumed throughout.

\section{Data and Methodology}\label{sec:sample}

\subsection{The AGN Sample}

The AGNs in this study were drawn from the MPA SDSS-DR4 type 2 AGN
catalog \citep{kauffmann03,brinchmann04}.  Type 2 AGNs are thought
to be the obscured counterparts of type 1 AGNs according to
unification models \citep[e.g.,][]{antonucci93,urry95}. The parent
sample was selected from SDSS emission-line galaxies at $z < 0.3$
based on diagnostic line ratios that distinguish AGN from stellar
ionizations \citep{bpt81}.  The four relevant lines (\hbeta,
\OIIIb, \halpha, and \NIIb ) were detected with signal to noise
ratios (S/N) $>3$.  We adopt the empirical criterion, log(\OIIIb
$/$\hbeta )$> 0.61/$(\{log(\NIIb $/$\halpha )$-0.05)+1.3$\}, which
separates AGNs from star-forming galaxies \citep{kauffmann03}. The
analysis is limited to \OIIIb\ luminosity \loiii\ in the range of
$10^{6.5}$--$10^{8.5} L_{\odot}$ which corresponds to bolometric
luminosities $L_{{\rm Bol}}\sim10^{43}$--$10^{45}$ erg s$^{-1}$
\citep[e.g.,][]{heckman04,reyes08,liu09}.  The lower bound is to
avoid faint-end incompleteness in a flux limited sample; It also
effectively excludes low ionization objects.  The upper bound is
to avoid substantial scattered-AGN contamination
\citep{zakamska06,liu09,greene09}, which makes population analysis
more complicated and uncertain.  Results on the \loiii\ $>10^{8.5}
L_{\odot}$ objects will be presented elsewhere.  The resulting
sample consists of 20,541 objects with a median redshift of 0.1.
These include 11,266 AGN-dominated objects with log(\OIIIb
$/$\hbeta )$> 0.61/$(\{log(\NIIb $/$\halpha )$-0.47)+1.19$\} above
the theoretical ``starburst limit'' \citep{kewley01} and 9,275
starburst-AGN composites\footnote{It is important to
include starburst-AGN composites in the analysis because (i) they
are a significant population; and (ii) otherwise the sample would
be biased against galaxies with high star formation and low AGN
activity.}.

\subsection{Bolometric Luminosity and Black Hole Mass}

We use \loiii\ (from the MPA catalog, uncorrected for
extinction\footnote{We adopt \loiii\ uncorrected for extinction
and the appropriate calibration rather than the
extinction-corrected method because the latter approach is
sensitive to uncertainties in the assumed reddening laws and
relies on over-simplified dust-screen models
\citep[e.g.,][]{reyes08}.}) to infer intrinsic AGN luminosity.
\loiii\ is observed to be correlated with broad-band continuum
luminosities in unobscured AGNs
\citep[e.g.][]{kauffmann03,reyes08}, despite having a significant
scatter \citep[e.g., $\sim$0.36 dex in log$L_{{\rm [O \,III]}}$ in
the correlation with $M_{2500}$][]{reyes08}.  We adopt the
conversion ${\rm log}(L_{{\rm Bol}}/L_{\odot}) = 0.99 \times {\rm
log}(L_{{\rm [O\,\, III]}}/L_{\odot}) + 3.5 (\pm 0.5)$ appropriate
for \loiii\ before extinction correction \citep{liu09}. We correct
\loiii\ for the contribution from star formation using an
empirical method based on an object's position on the \NII
/\halpha\ vs \OIII /\hbeta\ diagram following \citet{kauffmann09}.
SMBH mass is estimated ($M_{{\rm BH}}$) from stellar velocity
dispersion ($\sigma_{\ast}$), assuming the relation observed in
local inactive galaxies \citep[e.g.,][]{gebhardt00,ferrarese00}
calibrated by \citet{tremaine02}.  We adopt $\sigma_{\ast}$ from
the SDSS \texttt{specBS} pipeline \citep{SDSSDR6} and exclude
uncertain $\sigma_{\ast}$ measurements ($< 10$ or $> 600$ km
s$^{-1}$).

\subsection{Differentiated Specific Star Formation Rate}\label{subsec:ssfr}

The common emission-line SFR indicators cannot be directly applied
to luminous AGNs due to AGN contamination.  Continuum indices are
adopted such as D$_n(4000)$
\citep[e.g.,][]{kauffmann03,brinchmann04}.  To achieve higher
temporal resolution than D$_n(4000)$, we estimate SSFR averaged in
the past $\tau$ years (SSFR$_{\tau}$) from population synthesis
analysis \citep[e.g.,][]{bica88,cid05}.  Compared to SFR, SSFR is
less subject to absolute calibration uncertainties associated with
fiber effects and extinction \citep[e.g.,][]{brinchmann04}. We
measure SSFRs inside the SDSS 3\arcsec fibers (corresponding to
the central $\sim$1.5 kpc at typical redshift of the sample) to
quantify the bulge SF.

We adopt continuum models from the MPA DR4 archive
\citep{tremonti04}, and have tested them on a randomly drawn
subsample with our code \citep{liu09}.  The assumption is that the
bulk stellar population was built up in multiple starburst events
with ages smaller than the age of the universe at the galaxy
redshift.  To sample the whole history of a host galaxy, stellar
continuum is fit as a linear combination of ten instantaneous
starburst models (broadened with the measured $\sigma_{\ast}$)
with ages of 0.005, 0.025, 0.10, 0.29, 0.64, 0.90, 1.4, 2.5, 5.0,
and 11 Gyr \citep{bc03}, along with dust attenuation as an
additional parameter.  There are three sets of models with
different metallicities (0.2, 1.0, and 2.5 solar). We adopt the
metallicity with the smallest $\chi^2$ as the best estimate. Given
a best-fit model, SSFR$_{\tau}$ is calculated as the summed ratio
of the mass and age of the starburst templates that fall in
interval $\tau$, divided by the total mass contained in all
templates.  The model coefficients in the MPA DR4 archive are in
terms of luminosity\footnote{We caution that the SSFRs in
\citet{chen09} appear to be inferred using the luminosity
coefficients directly without the requisite mass-to-light
conversion.}, and need to be converted to {\it mass} weights using
the mass-to-light ratios of the starburst models from
\citet{bc03}.

\begin{figure*}
 \centering
 \includegraphics[width=82mm]{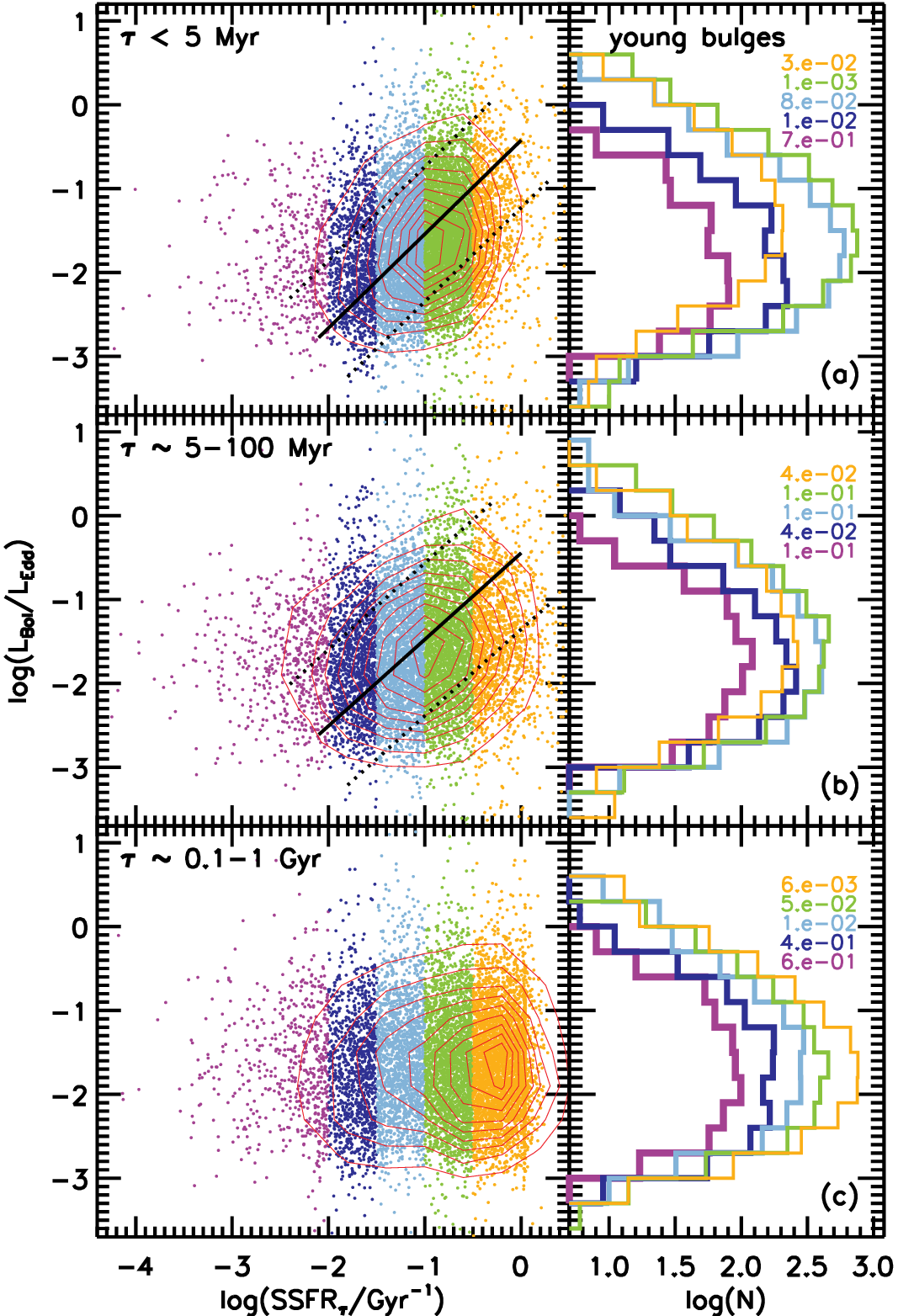}
 \includegraphics[width=82mm]{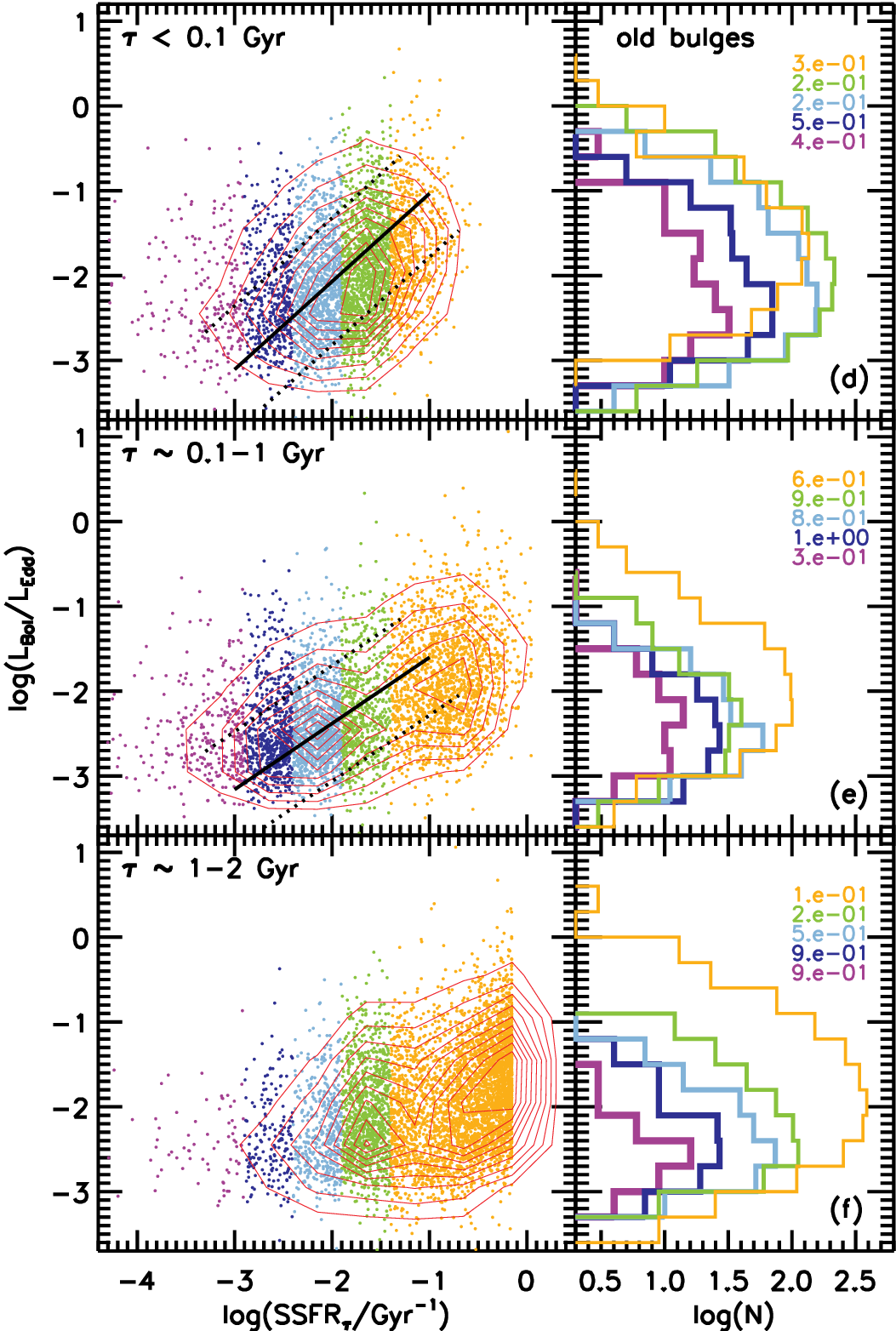}
 \caption{Differentiated specific star formation rate
    (\ssfr) versus Eddington ratio $\lambda$.
    The panels (a)--(c) are for young bulges (D$_n$(4000) $<1.6$)
    whereas the panels (d)--(f) are for old bulges (D$_n$(4000) $>1.6$).
    Look-back time intervals $\tau$ are labelled on each plot.
    Data and distribution curves are color coded in
    \ssfr\ (also discriminated with line thickness).
    The numbers (also color-coded in \ssfr ) are the
    Kolmogorov-Smirnov (KS) significance denoting the
    probability that a distribution is normal.
    Red contours represent ten
    evenly spaced levels in number density.  Black
    solid and dotted lines show the bisector linear regression
    fits (median and 1-$\sigma$ uncertainties; Table \ref{table:fit}).}
 \label{fig:ssfr}
\end{figure*}

We study the relation between AGN activity and recent bulge SF
over three characteristic time intervals.  First we divide the
sample into young and old bulges, because the distribution of mean
stellar age is bimodal \citep{kauffmann09}.  For ``young'' bulges
(D$_n$(4000) $< 1.6$)\footnote{Our results are independent of the
specific D$_n$(4000) division value chosen in the range of 1.6 to
1.8, but we adopt 1.6 so that the old bulges contain sufficient
objects for reliable statistics.}, the characteristic times are
$\tau \lesssim 5$ Myr (containing the age grid 0.005 Gyr), $\sim
5$--100 Myr (0.025 and 0.10 Gyr), and $\sim 0.1$--1 Gyr (0.29,
0.64, and 0.90 Gyr).  The large separations ($\sim$1-dex) between
these time intervals ensure the recovered SF at different epochs
to be well distinguished from one another
\citep[e.g.,][]{ocvirk06}.  Higher temporal resolution would cause
over interpretation due to degeneracies between models with
similar ages.  The timescales are characteristic of: 1--10 Myr for
massive star lifetimes (with masses $\gtrsim 10\, M_{\odot}$),
$<$0.1 Gyr for luminous AGN lifetime
\citep[e.g.,][]{yu02,hopkins09}, and 1--2 Gyr for lifetime of
lower-luminosity AGNs \citep[e.g.,][]{hopkins09} to which
D$_n(4000)$ is most sensitive to
\citep[e.g.,][]{bc03,kauffmann03}.  For ``old'' bulges
(D$_n$(4000) $> 1.6$), there is little current SF, and
SSFR$_{\tau}$ cannot be reliably estimated for $\tau \ll 0.1$ Gyr
(little mass allowed in the youngest starburst grids). The adopted
time intervals are $\tau \lesssim 0.1$ Gyr (containing the age
grids 0.005, 0.025, and 0.10 Gyr), $\sim 0.1$--1 Gyr (0.29, 0.64,
and 0.90 Gyr), and 1--2 Gyr (1.4 Gyr).

\subsection{Uncertainties and Selection Biases}

The uncertainty of $\lambda$ is dominated by the systematics of
$L_{{\rm Bol}}$ and $M_{{\rm BH}}$.  The quoted 0.5-dex 1-$\sigma$
uncertainty on $L_{{\rm Bol}}$ contains the intrinsic scatter of
the \loiii --$M_{2500}$ relation and the uncertainty of bolometric
correction in unobscured AGNs \citep{liu09}.   For $M_{{\rm BH}}$,
the uncertainty is propagated from $\sigma_{\ast}$ added in
quadrature with the 0.3-dex scatter of the $M_{{\rm BH}}$-$\sigma$
relation \citep{tremaine02}.  The total 1-$\sigma$ uncertainty of
$\lambda$ is $\sim 0.6$ dex.  We have tested effects of Malmquist
biases due to these scatters, and found that there is little bias
on $M_{{\rm BH}}$, since the $\sigma_{\ast}$ distribution is close
to log normal. On the other hand, there is a small yet detectable
bias on $L_{{\rm Bol}}$, such that the inferred $L_{{\rm Bol}}$ is
an overestimate (by $\lesssim$ 0.03 dex) of the {\it true} value
(assuming $L_{{\rm Bol}}$ is a prior whereas \loiii\ is a
posterior) given an observed bottom-heavy distribution of \loiii\ \citep[e.g.,][]{shen08}.

We have empirically tested the continuum-\ssfr\ approach using
12,090 SDSS star-forming galaxies \citep[e.g.,][]{brinchmann04}.
Our initial calibrations were carried out on high-S/N objects
(with median S/N $> 20$ pixel$^{-1}$ and S/N $>5$ in
$\sigma_{\ast}$). To constrain selection biases we tested low-S/N
objects using co-added spectra and found similar results. We have
verified that \ssfr\ correlate with the current SSFR$_{e}$ using
the MPA SFR$_{e}$ estimates from emission-line modeling
\citep{brinchmann04}.  SSFR$_{< 5\, {\rm Myr}}\approx$ SSFR$_{e}$
with a 1-$\sigma$ scatter of $\sim$0.25 dex in a bisector fit;
\ssfr\ with larger $\tau$ correlates with but does not equal to
the current SSFR$_{e}$, probably indicating varying SF as a
function of look-back time.  While these tests are limited to
star-forming galaxies, they do lend support to our general
approach, especially for young bulges.  We take the 0.25-dex
scatter as our fiducial estimate for the \ssfr\ systematic
uncertainty. This is comparable to the 0.3-dex estimate from
\citet{asari07} by comparing continuum-based SFRs against
H$\alpha$ SFRs for SDSS star-forming galaxies.  We have added
noise to spectra of a randomly selected subsample, and found that
typical errors of SSFRs due to noise in the continuum spectra are
no larger than $\sim0.2$ dex. We take $0.3$ dex as our fiducial
estimate for the total 1-$\sigma$ uncertainty for \ssfr\ .

\section{Results}\label{sec:corr}

Fig. \ref{fig:ssfr} shows Eddington ratio ($\lambda$) versus \ssfr
, and the $\lambda$ distribution (P(log$\lambda$)) as a function
of \ssfr\ . The young bulges (D$_n$(4000) $< 1.6$) have median
log(\loiii\ $/L_{\odot})=6.89$ (with a 1-$\sigma$ scatter of 0.39
dex) and log($M_{\ast}/M_{\odot}) = 10.88$ ($\pm$0.32), whereas
the old bulges (D$_n$(4000) $> 1.6$) have median log(\loiii\
$/L_{\odot})=6.73$ ($\pm$0.35) and log($M_{\ast}/M_{\odot}) =
11.13$ ($\pm$0.33).  For $\tau \leqslant \tau_{0}$ ($\tau_{0}
\sim$ 0.1 Gyr for young and $\sim 1$ Gyr for old bulges), both
populations appear to follow power laws.  The correlations are
strong: the Spearman probabilities for null correlation are close
to zero (Table 1, Col. 8); the relations flatten out when $\tau
\gtrsim \tau_{0}$ (Table 1, Col. 7).  The cutoff in Fig.
\ref{fig:ssfr}f at the high \ssfr\ end is artificial due to its
definition and the limited temporal resolution of the continuum
approach\footnote{The cutoff is reached if all mass is recovered
in the 1.4 Gyr grid. In practice this only happens for $\tau > $1
Gyr. For younger ages, the cutoff is never reached, because while
most of the light may go into young populations, it is always the
old populations which dominates the {\it total} mass (due to the
drastically different mass-to-light ratios of old and young
stars).}.

\subsection{The $\lambda$--${\rm SSFR}$ Relation}\label{subsec:scaling}

$\lambda$ of AGNs in young bulges correlate with their current
SSFRs (SSFR$_{\tau < 5 {\rm Myr}}$; Fig. \ref{fig:ssfr}a). We fit
the relation $\lambda = 10^b({\rm SSFR}/{\rm Gyr}^{-1})^k$, the
linear regression results of which are given in Table
\ref{table:fit} in. We adopt bisector fits appropriate for two
variables with no assumption of one depending on another.  The fit
for young bulges is log$\lambda = (-0.42\pm 0.01) +
(1.12\pm0.01)$log$({\rm SSFR}/{\rm Gyr}^{-1})$. The scatter
(1-$\sigma$ orthogonal of 0.54 dex) is dominated by observational
uncertainties.  Given any fixed \ssfr , P($\lambda$) is roughly
log normal, and its central value increases with increasing \ssfr
. While some of the Kolmogorov-Smirnov (KS) probabilities labelled
in Fig.\ref{fig:ssfr} are small, when we divide the sample into
finer bins in \ssfr\ the KS probabilities become larger.

In old bulges with measurable SSFRs, $\lambda$ also correlates
with recent SSFRs (SSFR$_{\tau < 0.1 {\rm Gyr}}$; Fig.
\ref{fig:ssfr}d).  The fit for old bulges is log$\lambda =
(0.00\pm 0.01) + (1.00\pm0.01)$log$({\rm SSFR}/{\rm Gyr}^{-1})$
(Table \ref{table:fit}).  Considering uncertainties, this is
similar to the relation in young bulges, although the median
$\lambda$ is $\sim 1$ dex smaller than in young bulges. Similarly,
given any fixed \ssfr , P($\lambda$) is close to log normal, and
the central $\lambda$ increases with increasing \ssfr .

\subsection{Correlation between AGN Activity and Recent Bulge Star Formation}

In young bulges $\lambda$ also correlates with SSFRs averaged over
the past $\sim 5$--100 Myr (Fig. \ref{fig:ssfr}b). The correlation
for $\tau \sim 5$--100 Myr is similar to $\tau\sim$ 5 Myr (Table
\ref{table:fit}).  For young bulges, the correlation flattens out
when $\tau \sim0.1$--1 Gyr (Fig. \ref{fig:ssfr}c): given a fixed
\ssfr , P($\lambda$) is still largely log normal, yet $\lambda$
depends little on \ssfr\ averaged over the past $\sim 0.1$--1 Gyr.
Figs. \ref{fig:ssfr}(e,f) show similar results in old bulges with
the main difference of having a larger $\tau_{0}$ ($\sim 1$ Gyr).
While there is still some correlation for $\tau \sim$ 1--2 Gyr
(Figs. \ref{fig:ssfr}f) partially due to the artificial cutoff,
the relation starts to flatten out (Spearman $\rho$ being smaller;
Table \ref{table:fit}).

\begin{table}
 \centering
 \begin{minipage}{100mm}
  \caption{Results of the fits ${\rm log}(\lambda) = b +
k\, {\rm log}({\rm SSFR}_{\tau}/{\rm Gyr}^{-1})$\label{table:fit}}
   \begin{tabular}{@{}cccccccc}
    \hline
    Sample &
    $\tau$/Gyr &
    $b$ &
    $k$ &
    $\sigma$ &
    N &
    $\rho$ &
    $P_{{\rm null}}$ \\
    (1) &
    (2) &
    (3) &
    (4) &
    (5) &
    (6) &
    (7) &
    (8) \\
    \hline
 young & $\leqslant 0.005$ &  $-0.42$  & $1.12$ & $0.54$ & 9903 & 0.23  & $<10^{-40}$  \\
 young & 0.005--0.1        &  $-0.45$  & $1.03$ & $0.63$ & 8420 & 0.15  & $<10^{-40}$  \\
 young & 0.1--1            &  $-0.81$  & $1.00$ & $0.68$ & 9951 & 0.01  & $0.5$        \\
 old   & $\leqslant 0.1$   & ~~$0.00$  & $1.00$ & $0.52$ & 3921 & 0.29  & $<10^{-40}$  \\
 old   & 0.1--1            &  $-0.82$  & $0.78$ & $0.54$ & 4302 & 0.29  & $<10^{-40}$  \\
 old   & 1--2$^{\ddag}$    &  $-0.37$  & $1.04$ & $0.48$ & 5351 & 0.16  & $10^{-16}$   \\
    \hline
   \end{tabular}
  \medskip\\
Col. (1): D$_n(4000) < 1.6$ for ``young'' whereas D$_n(4000) >
1.6$ for ``old'' bulges; Col. (2): look-back time $\tau$; Cols.
(3) \& (4): normalization and slope from the bisector fits with
1-$\sigma$ uncertainties of $\lesssim$0.01 dex; Col. (5): standard
deviation of the distribution. $\sigma$ is measured along the
orthogonal direction to the fit; Col. (6): number of objects in
each subsample.  To be included, a galaxy needs to contain mass in
at least one of the starburst grids in interval $\tau$; Cols. (7)
\& (8): Spearman correlation test coefficients. \ddag\ for Fig.
\ref{fig:ssfr}f, the fits are for 2499 objects with log$({\rm
SSFR}/{\rm Gyr}^{-1})< -1$, to minimize effects from the
artificial cutoff at the high SSFR end.
 \end{minipage}
\end{table}

\section{Discussion}\label{sec:discuss}

Our results imply similar relations between AGN activity and
recent SF in old and young bulges, despite different Eddington
ratios observed.  The current Eddington ratio correlates with the
most recent bulge SF: $\lambda \propto$ \ssfr\ estimated in a
certain look-back time interval $\tau$; it depends little on SF
happened before some threshold time $\tau_{0}$.  The main
difference between young and old bulges is on $\tau_{0}$: young
bulges have $\sim 0.1$ Gyr whereas old bulges have $\sim 1$ Gyr.

Recently \citet{kauffmann09} have studied the $\lambda$
distribution as a function of D$_n$(4000). These authors find that
P($\lambda$) is log normal in young bulges and exponential in old
bulges, and suggest two growth regimes.  While this could still be
the case, \citet{hopkins09} suggest that different $\lambda$
distributions may also arise from universal AGN light curves with
different quench times: AGNs in old bulges have apparent
exponential P($\lambda$) because accretion and star formation were
quenched in a series of earlier epochs \citep[see also e.g.,][]{shen09}.  Our results point to the
latter explanation.  Namely, the physical processes that link AGN
activity and bulge SF seem to work universally in young and old
bulges, albeit with different quenching/threshold times.

Using population synthesis analysis, we have found that when
grouped in terms of similar SF activity with ages less than
typical AGN lifetimes, BHs in old bulges also have log normal
P($\lambda$).  We have shown that for BHs in young bulges,
$\lambda$ does correlate with the most recent SF (Figs.
\ref{fig:ssfr}a,b).  The reason why \citet{kauffmann09} found
little correlation is most likely that the temporal resolution of
D$_n$(4000) is not enough to disintegrate the most recent SF in
young bulges.
The reciprocal SSFR$_{\tau}^{-1}$, which is the mass-weighted
stellar age averaged among the stellar populations formed over
$\tau$, can be estimated for different timescales $\tau$, whereas
D$_n$(4000) is most sensitive to $\tau = $ 1--2 Gyr.  If we assume
the observed sample consists of AGNs at different evolutionary
stages, the $\lambda$-SSFR results (Table \ref{table:fit}) imply
light curves $\lambda \propto t^{-(1.0-1.1)}$, which is noticeably
less steep than predictions $\lambda \propto t^{-(1.5-2.0)}$ from
models assuming extreme gas expel by AGNs
\citep[e.g.,][]{hopkins09}. This milder decay could suggest more
moderate self-regulated AGN and SF activity, although the
observational uncertainties in our results are significant.

Another implication is that AGN environmental studies should draw
control samples matched not only in the mean stellar age (as
usually implemented using D$_n$(4000)), but also in the most
recent bulge SF properties.  The effect would be particularly
important for AGNs in young bulges, as we have shown that their
SF-AGN-correlation timescale is usually much shorter ($\sim$ 0.1
Gyr) than that probed by the overall mean stellar age.

Our results are limited to optical AGNs with $\bar{z}\sim$0.1 and
$L_{{\rm Bol}}\sim10^{43}$--$10^{45}$ erg s$^{-1}$, and may not
apply to other AGN populations.  In principle the approach could
be extended to lower-luminosity AGNs (LINERs) and to
higher-luminosity AGNs and quasars, although selection
incompleteness and biases may be more severe. Multi-wavelength SFR
indicators that probe different SF timescales or phases may also
help elucidate the AGN-SF relation as a function of look-back
time.  The approach demonstrated based on population synthesis
analysis relies on statistical AGN samples with high-S/N spectra.
We look forward to the next generation of ground-based near-IR
multi-object spectrographs \citep[e.g.,][]{gunn09} that will make
an equivalent study possible at high redshifts to better
understand the coupled growths of stellar bulges and SMBHs in the
early universe.

\section*{Acknowledgments}

I am grateful to my thesis advisor M. Strauss for his generous
encouragement and for comments on the manuscript.  I also
acknowledge Y. Shen and J. Gunn for helpful discussion on
selection biases, and an anonymous referee for a careful and
useful report.
Funding for the SDSS and SDSS-II has been provided by the Alfred
P. Sloan Foundation, the Participating Institutions, the National
Science Foundation, the U.S. Department of Energy, the National
Aeronautics and Space Administration, the Japanese Monbukagakusho,
the Max Planck Society, and the Higher Education Funding Council
for England. The SDSS Web Site is http://www.sdss.org/.
%
%
Facilities: Sloan

{\small
\bibliography{msigmarefs}
}


\end{document}